\documentclass[conference]{IEEEtran}
\IEEEoverridecommandlockouts
\usepackage{cite}
\usepackage{amsmath,amssymb,amsfonts}
\usepackage{algorithmic}
\usepackage{graphicx}
\usepackage{textcomp}
\usepackage{xcolor}

\usepackage{cite}
\usepackage{amsmath,amssymb,amsfonts}
\usepackage{algorithmic}
\usepackage{graphicx}
\usepackage{subcaption}
\usepackage{wrapfig}
\usepackage{textcomp}
\usepackage{xcolor}
\usepackage{amsmath}

\usepackage{url}
\usepackage{tikz}

\usepackage{multirow}

\def\BibTeX{{\rm B\kern-.05em{\sc i\kern-.025em b}\kern-.08em
    T\kern-.1667em\lower.7ex\hbox{E}\kern-.125emX}}
\begin{document}

\author{\IEEEauthorblockN{ Bartosz Ku\'smierz}
\IEEEauthorblockA{\textit{$^1$ IOTA Foundation}\\ 10405 Berlin, Germany\\
$^2$\textit{Department of Theoretical Physics},\\ Wroclaw University of Science and Technology, Poland\\
bartosz.kusmierz@pwr.edu.pl}
\and
\IEEEauthorblockN{Roman Overko}
\IEEEauthorblockA{\textit{IOTA Foundation}\\ 10405 Berlin, Germany\\
roman.overko@iota.org}
}

\title{How centralized is decentralized? Comparison of wealth distribution in coins and tokens }

\maketitle

\begin{abstract}
Rapidly growing distributed ledger technologies (DLTs) have recently received attention among researchers in both industry and academia. While a lot of existing analysis (mainly) of the Bitcoin and Ethereum networks is available, the lack of measurements for other crypto projects is observed. 
This article addresses questions about tokenomics and wealth distributions in cryptocurrencies. We analyze the time-dependent statistical properties of top cryptocurrency holders for 14 different distributed ledger projects. The provided metrics include approximated Zipf coefficient, Shannon entropy, Gini coefficient, and Nakamoto coefficient. We show that there are quantitative differences between the coins (cryptocurrencies operating on their own independent network) and tokens (which operate on top of a smart contract platform). Presented results show that coins and tokens have different values of approximated Zipf coefficient and centralization levels. This work is relevant for DLTs as it might be useful in modeling and improving the committee selection process, especially in decentralized autonomous organizations (DAOs) and delegated proof of stake (DPoS) blockchains.

\end{abstract}


\section{Introduction}

The advent of Bitcoin~\cite{Bitcoin} has given rise to an increasing interest in distributed systems throughout the 2010s. The newly created space of cryptocurrencies has attracted many scientists, programmers, and business investors. Due to the complexity of Distributed Ledger Technologies (DLTs), their development requires expertise in many fields of science, including applied mathematics, cryptography, game theory, economics, peer-to-peer (p2p) networks, and coding theory. In the first years of DLTs, questions of technological nature received the most attention as the problems like consensus mechanism and peer-to-peer layer are at the core of any such technology. Unfortunately, questions about economics, cryptocurrencies  distributions and tokenomics took a back seat in academic research of cryptocurrencies and have not been sufficiently addressed (with a few notable exceptions).

This is unfortunate as the Bitcoin pseudo-anonymous account model allows for transaction transparency unprecedented in traditional financial systems, where almost all payments are private and highly sensitive. Furthermore, Bitcoin enabled new monetary models and deployed them on a global scale. Notably, the amount of Bitcoin currency units is capped at 21 million. However, due to some Bitcoin wallets being lost, as a result of negligence or human error, Bitcoin’s monetary policy is effectively deflationary. The monetary policy is not the only important factor cryptocurrency distribution. Even technology solutions like consensus mechanisms might influence cryptocurrency distribution. In this context, a comparison of Proof-of-Work (PoW) and Proof-of-Stake (PoS) consensus mechanisms is very informative. In PoW, newly created units of currency are rewarded to the specialized users, called miners, who have access to efficient Application-Specific Integrated Circuit (ASIC). PoW miners might hold a large number of cryptocurrency units; however, a large portion of mined rewards must be sold to cover expenses like electricity bills, rent, and amortization costs of ASIC machines.
In PoS systems, however, new tokens are rewarded to stakers who hold a large number of cryptocurrency units. Unlike PoW miners, PoS stakers do not experience high costs and are incentivized not to sell their rewards as doing so increases their revenue in the future. This illustrates that even supposedly monetary-agnostic technology solutions might influence tokenomics.

This paper partially addresses the questions of cryptocurrency tokenomics. We analyze the distribution of the top richest accounts in cryptocurrencies like Bitcoin, Ethereum, and selected ERC20 tokens. Our analysis involves in the data sets snapshotted at different dates with a given time interval. We use such data sets to measure different statistical metrics and analyze their evolution over time. Previous studies~\cite{com_sel,kondor2014dotherich,li2020comparison} showed that the distributions of the top richest balances might be modeled with Zipf's law. We expand on these results and study the time evolution of Zipf's law coefficient associated with such distributions. Notably, we analyze cryptocurrencies that, to our best knowledge, have never been analyzed before using similar methods. Next, we proceed with a thorough analysis of a series of centralization metrics like Shannon entropy, Gini index, and Nakamoto coefficient. These metrics are used to answer the main question addressed in this paper, which is  formulated as follows: {\it Are there any quantitative differences between top account balances in cryptocurrency ``coins'' and ``tokens''?} Therefore, the novelty of this work comprises the following two aspects: (i) studying quantitative differences between coins and tokens and (ii) examining cryptocurrencies whose analysis is missed in the literature.

A distinction between cryptocurrency {\it coins} and {\it tokens} was made in~\cite{zipf_mc}, where authors define coins as operating on their own independent ledger/network and tokens as operating on top of a coin network (typically smart contract platforms as Ethereum or Cardano). For the purposes of this paper, we use the same definitions.

This research might be particularly interesting for DLTs, where a group of top cryptocurrency holders fulfills a special role. Examples include Decentralized Autonomous Organizations (DAOs) in which a committee of top token holders is responsible for DAO governance or treasury management. Other examples are Delegated Proof-of-Stake (DPoS) blockchains, where a relatively small committee of block validators issues ledger updates or distributed random number generators based on the threshold signature scheme. Since our research is focused on a relatively small group of top token holders, it can be directly applied to modeling the aforementioned examples. This is also reasonable as the typical size of the threshold signature committee is limited by the message complexity (up to $50$---$100$ nodes). Our research might help to improve the committee selection process as we provide a range of parameters of Zipf's law coefficient, which might be used as a model of cryptocurrency distribution.

The structure of the paper is as follows. In the next section, we discuss related work and introduce the methods and tools used in this paper. Section~\ref{sec:results} is devoted to the presentation and analysis of the results. In the last section, we conclude our findings and discuss future research.

\section{Related work and methodology}

One of the first works analyzing Bitcoin using network characteristics over time (as well as the wealth statistics and the temporal patterns of transactions) is presented in~\cite{kondor2014dotherich}. The authors showed that the wealth of top Bitcoin holders grows faster than the wealth of low balance accounts---this phenomenon is well known as \emph{preferential attachment}, and it plays an essential role in the formation of the wealth distribution. Additionally, the Gini coefficients were also computed to measure wealth inequality. Analysis of the Bitcoin network from a perspective of mining pools can be found in~\cite{wang2020measurement}, in which the authors studied how characteristics of mining pools such as computing power, hash rate, mining revenue, transaction collection strategies, and block size affect the security of the network, transaction delays, and fees. Their measurement results showed that more than $50\%$ of the blocks were created by the top 5 mining pools, which may raise security and centralization concerns for the Bitcoin network.

The results of similar research to our work were presented in~\cite{lin2021measuring}. Specifically, the authors provided their analysis using three different metrics (Gini coefficient, Shannon entropy, and Nakamoto coefficient) and their evolution over time. It was found that the degree of decentralization in Bitcoin is higher and more volatile, while the degree of decentralization in Ethereum is smaller and more stable. Jensen et al.~\cite{jensen2021decentralized} analyzed decentralization of governance token distribution in four decentralized finance (DeFi) applications on the Ethereum blockchain using Gini and Nakamoto coefficients. Their results indicated that the token distributions for all four DeFi applications are characterized by high Gini coefficients.  Similar methods were used in~\cite{li2020comparison}, where PoW and PoS-based cryptocurrencies were compared. The authors analyzed the decentralization of Bitcoin and Steem using Shannon entropy. However, their analysis was limited to only two cryptocurrencies and did not account for time changes in the distribution.

The authors of~\cite{zipf_mc} constructed a growth model for a market capitalization of cryptocurrencies using Gibrat's law. They also pointed out that crypto coins (which operate on their own independent DLT network) and crypto tokens (which operate on top of another coin platform) follow Zipf's law for their capitalization. Although, the parameter of Zipf's law is quantitatively different for coins and tokens.

A large part of research in such literature has been devoted to the analysis of the properties of cryptocurrencies networks from a graph perspective where each transaction is represented by a link in a graph of addresses~\cite{weili2020traveling,lehnberg2018valuing,somin2018network,somin2020erc20,friedhelm2019measuring,li2020comparison,lin2021measuring}. For example, Lehnberg~\cite{lehnberg2018valuing} aimed to determine whether the relationships between the users of ERC20 token networks and their valuations follow  Metcalfe’s law. It was found that only two tokens of 50 seem to obey Metcalfe’s law, while the rest follow a linear or sub-linear law.

\subsection{Data collection and cryptocurrency wallets}\label{sec:data}
Many blockchain networks do not store balances associated with addresses; however, balances can be calculated from the sum of sent and received assets (i.e., coins or tokens) for each address. In this work, we use public blockchain datasets available on Google BigQuery\footnote{\url{https://bigquery.cloud.google.com/dataset/bigquery-public-data}}. Starting from the genesis block of a particular blockchain, we collected historical transaction/transfer data until January 16, 2022, inclusive. In the case of some ERC20 tokens, the first few weeks of the data had been skipped before the analysis due to the insufficient number of transactions. This early period could have been devoted to testing or marketing. Including these data could bring artifacts\footnote{For example, there were no token transfers for the first few weeks after the inception of the Tether (USDT) smart contract.}

It is important to note that the data presented in this paper does not represent the wealth of individual cryptocurrency owners but rather the wealth distribution among the cryptocurrency wallets. Cryptocurrency wallets are not unique to a user, and one user can be in possession of multiple such wallets. The true identity of addresses owners is hard to establish and might be problematic even for entities that have access to the cryptocurrency exchanges' data. Aware of these limits, we find the distribution of wealth in the richest cryptocurrency wallets to be still interesting as the basis of identity systems in PoS systems and DAOs are wallets (see section \ref{sec:samplesize}).

\subsection{Sample size $N$}\label{sec:samplesize}

In this paper, we analyze the properties of the empirical distribution function for the top $N$ richest accounts. 
Such empirical distributions are discrete, and the value of the $i$-th  entry is the ratio of the $i$-th richest account balance to the sum of $N$ richest account balances. We focus on a relatively small sample size $N$ between $30$---$100$. These numbers might seem arbitrary and small, especially in the face of thousands and tens of thousands of cryptocurrency users. However, this interval is interesting for applications in DPoS DLTs based on Byzantine Fault Tolerance (BFT)~\cite{HB,pBFT,mirBFT} consensus mechanisms.    

\begin{figure}[ht]
    \centering
    \includegraphics[width=\columnwidth]{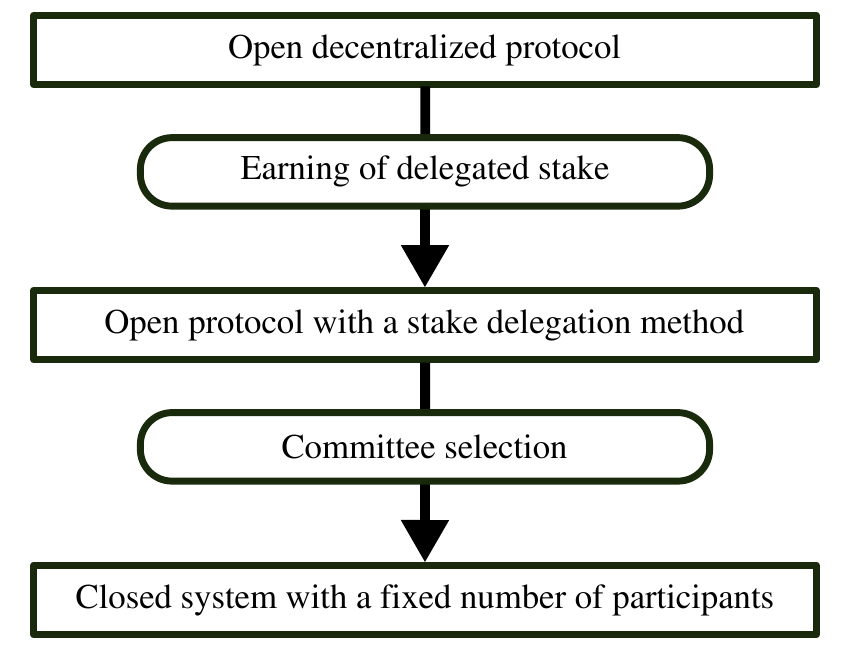}
    \caption{Process of finding fixed-size closed committee in open permissionless systems with stake delegation. }
    \label{fig:open-closed}
\end{figure}

In the most standard versions of  BFT consensus mechanisms, the number of consensus participants is fixed, and their identities are known. Such systems are permissioned and not suitable for direct use in fully permissionless DLTs. However, an interesting modification to BFT design is used by a series of DPoS blockchains. These projects use an intermediary step between open and permissionless networks. Any user is allowed to set up a node and collect tokens, but only the most reliable nodes with the most stake contribute to the consensus directly. An illustration of the procedure of establishing consensus based on the fixed-size closed committee in open and permissionless systems is depicted in Fig.~\ref{fig:open-closed}.

An example of a DPoS protocol is EOS\footnote{\url{https://eos.io/}}, in which blocks are produced by a committee of 21 validators who collectively sign the new blocks using an asynchronous version of the BFT consensus mechanism~\cite{EOS}. Committee members are selected and periodically rotated based on the amount of stake which was delegated to them by other network users. Other examples include Lisk\footnote{\url{https://lisk.com/}}, which uses $101$ nodes, and Internet Computer (ICP)\footnote{\url{https://dfinity.org/}} with the 101 nodes in the Neural Nervous System (ICP's main blockchain). In theory, the size of block producing committee can be unbounded. However, in practice, the procedure of signing new  blocks is limited by the bandwidth---the number of messages exchanged among the committee members grows like a square of its size. In the most practical applications, the block-producing committee has from $10$ to $50$ committee members. Some of the multi-blockchain protocols might afford to use up to 100 nodes in some parts of their protocol; however, in these cases, the block production time suffers. These limits explain our interest in a relatively small interval of the sample size of $N$, namely $30$---$100$, which might be used to model the token distribution among block validators or improve the committee selection process.

\subsection{Zipf's law} 

The Zipf's distribution is a discrete distribution commonly found in physics and social sciences. It was empirically confirmed that the Zipf's law describes a variety of effects in quantitative linguistics~\cite{Zipf+2013}, the study of country and city population~\cite{zipf_cities}, and web sites references and other effects. Perhaps the most relevant for this paper are applications of the Zipf's law in modeling the wealth in societies~\cite{wealth_pareto}, and the distribution of token holders in various cryptocurrencies~\cite{com_sel,kondor2014dotherich,li2020comparison}.

Zipf's distribution $f(k; N, s)$ is indexed by two parameters: (i) an integer number $N$, which is the number of elements, and (ii) a real-valued parameter $s$---the exponent of the distribution. The value of the $k$-th element satisfies
\begin{equation}\label{eq:zipf}
    f(k; N, s) = H(s, N)^{-1} k^{-s},
\end{equation}
where $H(s,N) = \sum_{i=1}^{N} i^{-s}$ is the normalization factor.

\subsection{Centralization metrics} \label{sec:def}

In this paper, we discuss multiple centralization metrics like Gini and Nakamoto coefficient. We want to stress again that analyzed sample size $N$ is relatively small and can significantly influence values of these metrics due to the normalization factor $H(s,N)$ in Eq.~\eqref{eq:zipf}. 

Analyzed metrics are denoted using variables labeled by an integer index $i$ such that $1 \leq i \leq N$. The percentage of assets (i.e., coins or tokens) held by the $i$-th user is denoted by $x_i$ such that $x_i$ are ordered and normalized to one, i.e., $x_1 \geq x_2 \geq ... \geq x_N$ and $\sum_{i=1}^N x_i = 1$.

\subsubsection{Entropy}
Entropy (Shannon entropy) is a centralization metric commonly used in physics, especially in quantum information theory (Von Neumann entropy), thermodynamics, and statistical physics. Entropy $S$ is defined as 
\begin{equation}\label{eq:entropy}
    S = -\sum^N_{i=1} x_i \log(x_i).
\end{equation}

In the general case, the values of entropy are unbounded. However, when the number of available states $N$ is fixed, then the entropy takes values from an interval $[0, \log(N)]$. Maximal centralization corresponds to the entropy equal to zero, and decentralization grows with entropy.

\subsubsection{Gini Coefficient}

The Gini coefficient $G$ is an inequality measure widely used in economics and social statistics:
\begin{equation}\label{eq:gini}
    G =\frac{1}{2N} {\displaystyle{\sum_{i=1}^N \sum_{j=1}^N \left| x_i - x_j \right|}}.     
\end{equation}
Gini index takes values from 0 (complete decentralization of wealth) to 1 (absolute centralization).

\subsubsection{Nakamoto Coefficient}

The Nakamoto coefficient $K$ is a relatively new centralization metric, popular in the analysis of cryptocurrencies. It is defined as
\begin{equation}\label{eq:nakamoto}
    {K} = \min \left \{n \in \mathbb{N}: \sum_{i=1}^{{n}} x_i > \frac{1}{2} \sum_{i=1}^{{N}} x_i  \right \},   
\end{equation}
which is the minimal number of actors who control more than half of the network resources. It was originally introduced to assess the feasibility of a $51\%$ attack on the Bitcoin network.

\section{Results}\label{sec:results}

\begin{figure}[!ht]
    \centering
    \includegraphics[width=0.45\textwidth]{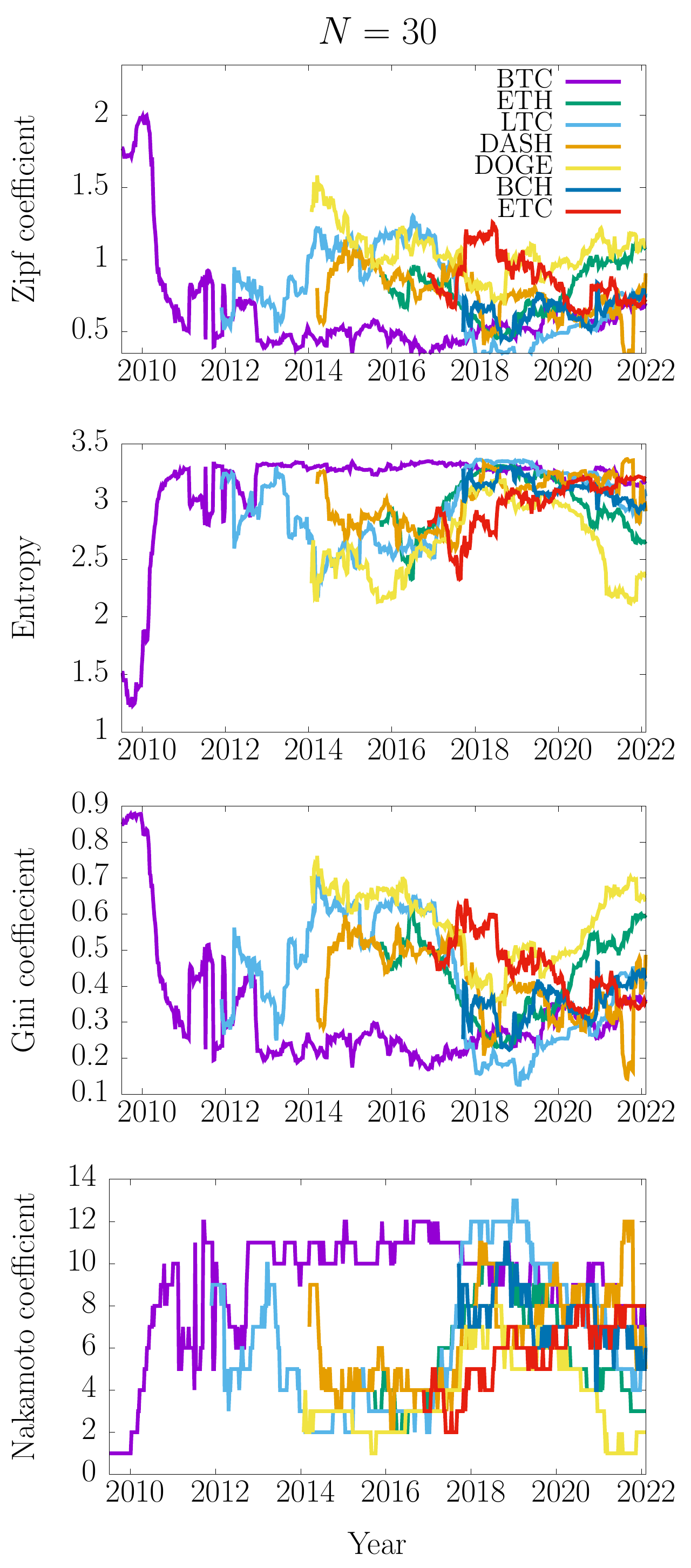}
    \caption{Time evolution of statistical metrics (Zipf's law coefficient, Shannon entropy, Gini coefficient, and Nakamoto coefficient) for selected cryptocurrency coins (sample size $N=30$).}\label{c30}
\end{figure}

\begin{figure}[!ht]
    \centering
    \includegraphics[width=0.45\textwidth]{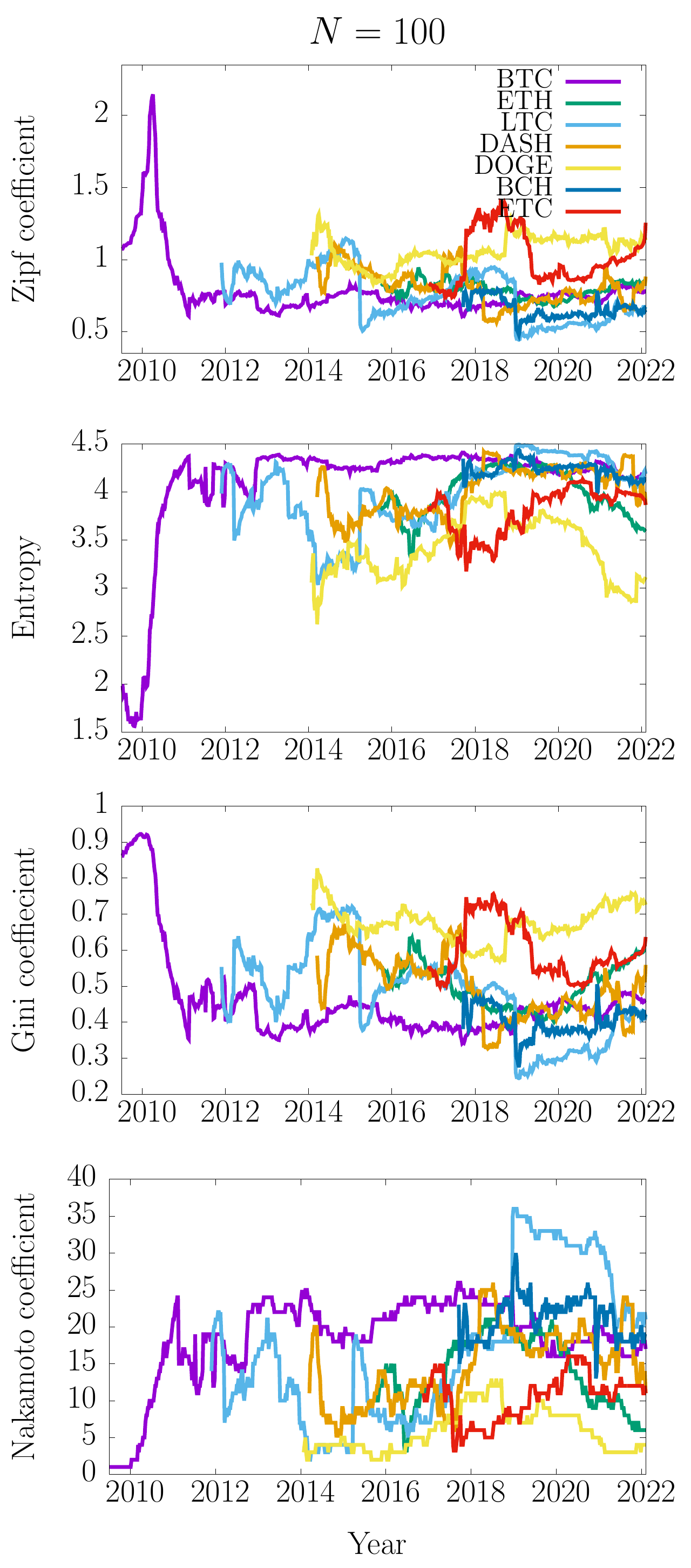}
    \caption{Time evolution of statistical metrics (Zipf's law coefficient, Shannon entropy, Gini coefficient, and Nakamoto coefficient) for selected cryptocurrency coins (sample size $N=100$). }\label{c100}
\end{figure}

\begin{figure}[!ht]
    \centering
    \includegraphics[width=0.45\textwidth]{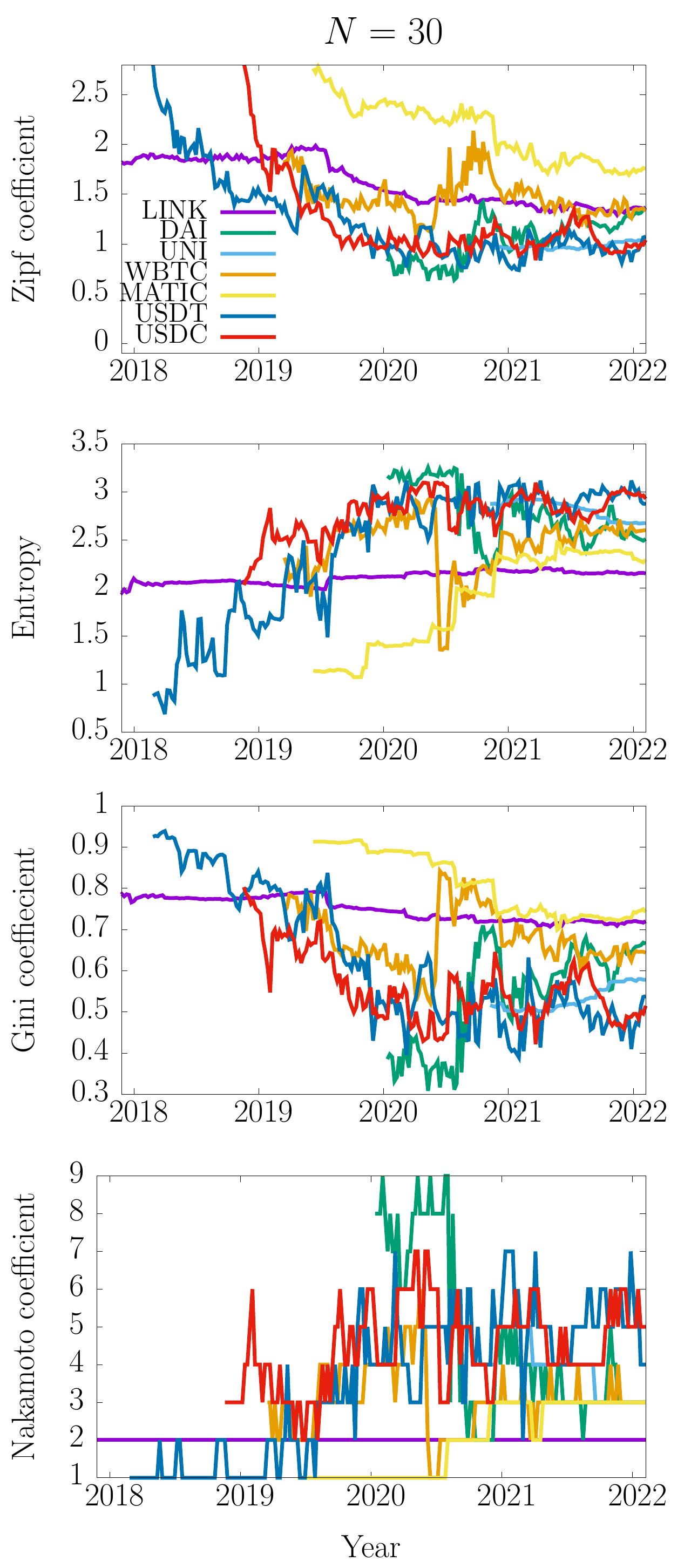}
    \caption{Time evolution of statistical metrics (Zipf's law coefficient, Shannon entropy, Gini coefficient, and Nakamoto coefficient) for selected ERC20 tokens (sample size $N=30$).}\label{t30}
\end{figure}

\begin{figure}[!ht]
    \centering
    \includegraphics[width=0.45\textwidth]{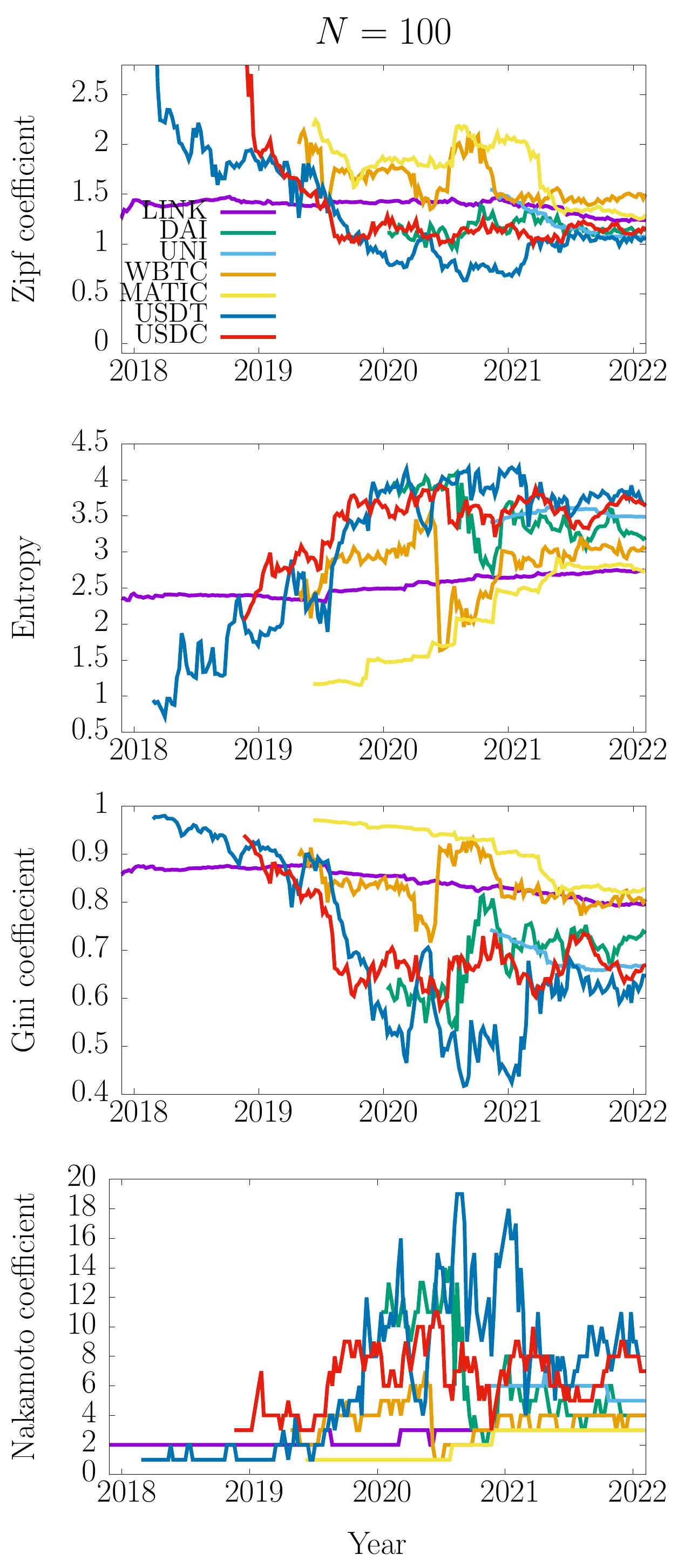}
    \caption{Time evolution of statistical metrics (Zipf's law coefficient, Shannon entropy, Gini coefficient, and Nakamoto coefficient) for selected ERC20 tokens (sample size $N=100$). }\label{t100}
\end{figure}

A series of plots that show the evolution of statistical metrics for the top accounts are given in Figs.~\ref{c30}-\ref{t100}. We present the data for two sample sizes $N=30$ and $N=100$. Cryptocurrencies are divided into two groups. The first group consists of coins, which include Bitcoin (BTC), Ethereum (ETH), Litecoin (LTC), DASH, Dogecoin (DOGE), Bitcoin Cash (BCH), and Ethereum Classic (ETC). The second group includes ERC20 tokens on the Ethereum chain: Chainlink (LINK), DAI, Uniswap (UNI), Wrapped Bitcoin (WBTC), Polygon (MATIC), Tether (USDT), and USD Coin (USDC). Note that BCH and ETC are ``forks’’ of BTC and ETH, respectively. This means that BCH and BTC (ETC and ETH) share ledger values before the fork\footnote{The ledgers of BCH and BTC split on August 1, 2017. The split date for ETC and ETH is July 20, 2016.}. To avoid data duplication, we present the data for forked coins only after the split date.

Each figure contains four subplots with (1) approximated Zipf's law coefficient, (2) Shannon entropy, (3) Gini coefficient, and (4) Nakamoto coefficient. The Zipf's law coefficients were obtained using linear regression on the double logarithmic scale, while the remaining metrics are computed using definitions from Section~\ref{sec:def}. Plots of time evolution were supplemented with the tables of average values of four analyzed quantities (see Tables~\ref{tz}-\ref{tn}).

\begin{table}[ht] \renewcommand{\arraystretch}{1.1}
    \centering
    \begin{tabular}{ |c|| c| c|c| }
        \hline
        Name	&	Top 30	&	Top 50	&	Top 100	\\	\hline \hline
        \multicolumn{4}{|c|}{Coins}\\\hline
        BTC	&	0.656	&	0.729	&	0.787	\\	\hline
        ETH	&	0.774	&	0.784	&	0.792	\\ \hline
        LTC	&	0.759	&	0.732	&	0.751	\\	\hline
        DASH	&	0.760	&	0.782	&	0.810	\\	\hline
        DOGE	&	1.053	&	1.068	&	1.063	\\	\hline
        BCH	&	0.638	&	0.631	&	0.658	\\	\hline
        ETC	&	0.865	&	0.953	&	1.011	\\	\hline 
        \hline
        \multicolumn{4}{|c|}{Tokens (ERC 20)}\\\hline
        LINK	&	1.622	&	1.504	&	1.386	\\	\hline
        DAI	&	1.044	&	1.100	&	1.155	\\	\hline
        UNI	&	0.981	&	1.033	&	1.242	\\	\hline
        WBTC	&	1.032	&	1.245	&	1.458	\\	\hline
        MATIC	&	2.151	&	2.008	&	1.808	\\	\hline
        USDT	&	1.131	&	1.170	&	1.247	\\	\hline
        USDC	&	1.180	&	1.256	&	1.322	\\	\hline
    \end{tabular}
    \caption{Average Zipf's law coefficient of the token distribution for the top 30, 50, and 100 token holders in selected cryptocurrencies.}\label{tz}
\end{table}

\begin{table}[ht]\renewcommand{\arraystretch}{1.1}
    \centering
    \begin{tabular}{	|c||	c|	c|c|	}	\hline														
        Name	&	Top	30	&	Top	50	&	Top	100	\\	\hline	\hline					
        \multicolumn{4}{|c|}{Cryptocurrencies}\\\hline							
	        BTC	&	3.104	&	3.514	&	4.056	\\	\hline					
            ETH	&	2.994	&	3.422	&	4.003	\\	\hline
	        LTC	&	2.960	&	3.401	&	3.990	\\	\hline								
	        DASH	&	3.017	&	3.445	&	4.012	\\	\hline								
            DOGE	&	2.654	&	2.986	&	3.427	\\	\hline								
			BCH	&	3.118	&	3.597	&	4.231	\\	\hline								
	        ETC	&	2.981	&	3.344	&	3.806	\\	\hline						
        \hline																	
        \multicolumn{4}{|c|}{Tokens	(ERC	20)}\\\hline															
	        LINK	&	2.108	&	2.288	&	2.521	\\	\hline								
        	DAI	&	2.787	&	3.102	&	3.487	\\	\hline									
        	UNI	&	2.821	&	3.169	&	3.526	\\	\hline								
        	WBTC	&	2.472	&	2.655	&	2.838	\\	\hline								
        	MATIC	&	1.850	&	1.931	&	2.050	\\	\hline								
							
        	USDT	&	2.382	&	2.667	&	3.032	\\	\hline								
        	USDC	&	2.731	&	3.012	&	3.365	\\	\hline								
    \end{tabular}											
	\caption{Average	Shannon entropy of the token distribution for the top 30, 50, and 100 token holders in selected cryptocurrencies.}  \label{te}
\end{table}

\begin{table}[ht]\renewcommand{\arraystretch}{1.1}																
    \centering
    \begin{tabular}{	|c||	c|	c|c|	}													
	    \hline																
        Name	&	Top	30	&	Top	50	&	Top	100	\\	\hline	\hline					
        \multicolumn{4}{|c|}{Coins}\\\hline											
        BTC	&	0.329	&	0.392	&	0.466	\\	\hline												
	    ETH	&	0.423	&	0.460	&	0.499	\\	\hline																						    
    	LTC	&	0.414	&	0.429	&	0.475	\\	\hline								
    	DASH	&	0.412	&	0.452	&	0.498	\\	\hline								
    	DOGE	&	0.571	&	0.623	&	0.674	\\	\hline							
    	BCH	&	0.350	&	0.367	&	0.402	\\	\hline									
    	ETC	&	0.451	&	0.523	&	0.597	\\	\hline								
        \hline																	
        \multicolumn{4}{|c|}{Tokens	(ERC	20)}\\\hline															
		LINK	&	0.749	&	0.799	&	0.846	\\	\hline								
    	DAI	&	0.538	&	0.607	&	0.685	\\	\hline								
    	UNI	&	0.531	&	0.587	&	0.688	\\	\hline								
    	WBTC	&	0.667	&	0.745	&	0.829	\\	\hline								
	    MATIC	&	0.813	&	0.868	&	0.907	\\	\hline								
							
    	USDT	&	0.633	&	0.670	&	0.711	\\	\hline								
    	USDC	&	0.568	&	0.637	&	0.711	\\	\hline								
	\end{tabular}																
	\caption{Average	Gini coefficient of the token distribution for the top 30, 50, and 100 token holders in selected cryptocurrencies.}  \label{tg}
\end{table}

\begin{table}[ht]\renewcommand{\arraystretch}{1.1}
    \centering
    \begin{tabular}{	|c||	c|	c|c|	}							
        \hline																
        Name	&	Top	30	&	Top	50	&	Top	100	\\	\hline	\hline					
        \multicolumn{4}{|c|}{Cryptocurrencies}\\\hline																	
		BTC	&	8.778	&	12.060	&	18.129	\\	\hline	
	    ETH	&	5.911	&	8.536	&	13.982	\\	\hline
	    LTC	&	6.399	&	10.170	&	16.940	\\	\hline					
	    DASH	&	6.440	&	9.125	&	14.594	\\	\hline								
	    DOGE	&	3.771	&	4.567	&	5.943	\\	\hline									
	    BCH	&	7.528	&	11.858	&	21.210	\\	\hline								
	    ETC	&	5.725	&	7.406	&	10.380	\\	\hline	\hline																	
        \multicolumn{4}{|c|}{Tokens	(ERC	20)}\\\hline															
	    LINK	&	2.000	&	2.049	&	2.469	\\	\hline								
    	DAI	&	4.661	&	5.536	&	6.723	\\	\hline								
    	UNI	&	3.899	&	4.841	&	5.768	\\	\hline								
    	WBTC	&	3.521	&	3.877	&	4.196	\\	\hline

    	    MATIC	&	1.992	&	2.014	&	2.021	\\	\hline								
    	USDT	&	3.595	&	4.512	&	6.516	\\	\hline								
    	USDC	&	4.385	&	5.224	&	6.402	\\	\hline					
    \end{tabular}
	\caption{Average	Nakamoto coefficient	of the token distribution for the top 30, 50, and 100 token holders in selected cryptocurrencies.	} \label{tn}
\end{table}													

A cursory analysis of the time-dependent data reveals that, in an initial period (just after inception), cryptocurrencies experience a volatile evolution, with rapid changes in analyzed quantities. The length of this initial period varies and can take from a couple of months to a couple of years in the case of the oldest coins. After the initial period ends, metrics evolve less dynamically, and their evolution resembles a random walk around an average value. A great example of this behavior is the data for Bitcoin. Bitcoin’s metrics show the greatest volatility in a period of approximately two years after the genesis block (see Fig.~\ref{c30} and Fig.~\ref{c100}). Curiously, the extreme values of analyzed quantities coincide with the date of the famous ``Bitcoin pizza transaction’’ (May 22, 2010)\footnote{The ``Bitcoin pizza transaction’’ is a Bitcoin transfer made by Laszlo Hanyecz who spent 10,000 Bitcoins at a local pizza restaurant to buy two pizzas.}. The ``Bitcoin pizza transaction’’ might be interpreted as Bitcoin leaving its ``infancy'', experimental stage, and slowly being adopted as a payment method. Before May 2010, Bitcoin was rarely exchanged among users and was not used as payment for commodities.

The data for ERC20 tokens were collected for a significantly shorter time period than coins (4 years vs. 12 years). This is expected since ERC20 tokens are relatively new cryptocurrencies deployed on the Ethereum blockchain. Moreover, due to the more speculative nature, ERC20 have typically much smaller market capitalization and less active addresses than coins.

The biggest outlier in the group of ERC20 tokens is Chainlink (LINK). The Chainlink token distribution is very concentrated on the whole time interval---with the Gini coefficient around $0.8$ and a surprisingly small Nakamoto coefficient of $2$. This is a result of an extreme number of tokens in the richest address dubbed ``Chainlink: Node Operators'' (even up to $35\%$ of the whole token supply)\footnote{See Ethereum address: 0x98c63b7b319dfbdf3d811530f2ab9dfe4983af9d}. Furthermore, two next richest addresses, which both belong to the team who develops the Chainlink token, have approximately $5\%$ of the total supply, which also adds to the extremely high centralization. Other tokens also experienced extremely high centralization (Tether, USD Coin), however, only Chainlink’s distribution is so concentrated on the whole analyzed interval.

Let us now focus on the main question of this paper, which is what the differences are between the statistical metrics of cryptocurrency coins and ERC20 tokens. We note that a similar, although not the same problem, was examined in~\cite{zipf_mc}, where the authors analyzed the market capitalization of cryptocurrencies. The authors found that capitalization of both coins and tokens follows qualitatively the same power law. However, there are quantitative differences; namely, the estimated Zipf's law coefficient for coins is between $0.5$---$0.7$ and for tokens, this value is closer to $1.0$---$1.3$.

In this paper, we address the question of quantitative differences in the token distributions of coins and tokens. Tables~\ref{tz}-\ref{tn} show that the analyzed statistical quantities for tokens and coins take different values. For example, the Zipf's law coefficient of different coins, is for the most part, confined within the range of $0.4$---$1.25$, while for ERC20 tokens, this range is $0.7$---$2.25$ (see the top panel in Figs.~\ref{c30}-\ref{t100}). Furthermore, coins are typically less concentrated---the Gini coefficient is in the range $0.1$---$0.7$, and the Nakamoto coefficient varies within $10$---$25$ (for sample size $N=100$). For tokens, the Gini coefficient is around $0.5$---$0.9$, and the Nakamoto coefficient varies within $2$-$8$ ($N=100$). Nevertheless, it is worth stressing that the Gini coefficient for the ERC20 tokens has a declining tendency, and it might reach levels similar to coins at the same maturity level. One illustration of a tendency for the decreasing centralization are stablecoins\footnote{Stablecoins are cryptocurrencies whose value is tied to a physical commodity like the US dollar, Euro, or gold.}  Tether and USD Coin. Both tokens start with extremely centralized token distribution as indicated by the Gini coefficient close to $1$, but, as time passes, centralization significantly decreases.

Very low values of the Nakamoto coefficients of ERC20 tokens need to be taken into account while designing technological solutions. Especially if a given ERC20 token is an underlying Sybil protection mechanism of a DAO. Some parts of DAOs with a small Nakamoto coefficient could be paralyzed or even hijacked by attacking only a tiny number of nodes. To give a practical example: one popular feature in DAOs is a distributed Random Number Generator (dRNG). Some of the dRNG designs~\cite{randHH} are based on threshold signatures, where a random number is produced by the committee of nodes. If the collective signature is composed of partial signatures, weighted by the amount of held tokens, then a small Nakamoto coefficient might threaten the liveness of the dRNG. To address such a problem, DAO should adapt dRNG schemes where all of the partial signatures have the same weight or use a completely different dRNG scheme. For example, based on Verifiable Delay Functions (VDFs)\footnote{VDFs can only be computed sequentially and cannot benefit from parallel computation~\cite{wes,pietrz}. Furthermore VDFs provide quickly verifiable proof of performed calculations.}.

\section{Summary}
In this article, we analyzed the wealth distribution of the richest addresses in various cryptocurrencies. This included the time evolution of statistical metrics like the approximated Zipf's law coefficient, Shannon entropy, Gini coefficient, and Nakamoto coefficient, along with their average values. It was shown that coins and ERC20 tokens have quantitatively different distributions of wealth. In particular, the values of approximated Zipf's law coefficient for coins are $0.4$---$1.25$ and $0.7$---$2.25$ for ERC20 tokens. Differences between the two groups were also apparent during the study of wealth centralization. It was observed that tokens are, in general, much more centralized than coins with higher Gini coefficients and smaller Nakamoto coefficients.

This research might be of particular interest to DAOs and DPoS-based blockchains which rely on some form of a committee of the richest token holders. Presented values of statistical metrics like approximated Zipf coefficient or Nakamoto coefficient might help to model a committee selection process and make it more secure.
Future work will evolve in two directions. Firstly, we are working on incorporating more metrics, analyzing more tokens, and considering more different sample size values. Secondly, the main findings of this work will be used to model a committee selection process.

\bibliographystyle{abbrv}
\bibliography{refs}

\end{document}